\documentclass[twocolumn]{aastex631}
\usepackage{CJK}
\usepackage{framed}
\usepackage{multirow}
\usepackage{makecell}
\usepackage{amsmath, bm}
\usepackage{graphicx}
\usepackage{amssymb}
\usepackage{amsmath}
\usepackage{times}
\usepackage{subfigure}
\usepackage[T1]{fontenc}

\newcommand{\Msun}{\>{\rm M_\odot}}

\newcommand{\Mag}{\>{\rm mag}}

\newcommand{\Mpch}{\>h^{-1}{\rm {Mpc}}}
\newcommand{\Mpc}{{\rm \ {Mpc}}}

\newcommand{\kpc}{{\rm \ {kpc}}}

\def \MF{{\rm M}_{\rm 500}}

\received{XXX}
\revised{YYY}
\accepted{ZZZ}
\submitjournal{ApJ}

\begin{document}
\begin{CJK*}{UTF8}{gkai}

\shorttitle{Diffuse light in simulations}
\shortauthors{Tang et al.}

\title[Diffuse light in simulations]{Mock Observations: Formation and Evolution of diffuse light in Galaxy Groups and Clusters in the IllustrisTNG Simulations}

\author[0000-0001-6395-2808]{Lin Tang}
\affiliation{School of Physics and Astronomy, China West Normal University, ShiDa Road 1, 637002, Nanchong, China}
\affiliation{CSST Science Center for the Guangdong-Hongkong-Macau Greater Bay Area, DaXue Road 2, 519082, Zhuhai, China}

\author[0000-0003-2204-2474]{Weipeng Lin}
\affiliation{School of Physics and Astronomy, Sun Yat-sen University, DaXue Road 2, 519082, Zhuhai, China}
\affiliation{CSST Science Center for the Guangdong-Hongkong-Macau Greater Bay Area, DaXue Road 2, 519082, Zhuhai, China}

\author[0000-0002-1512-5653]{Yang Wang}
\affiliation{Department of Mathematics and Theories, Peng Cheng Laboratory, No. 2, Xingke 1st Street, Nanshan District, Shenzhen, 518000, China}
\affiliation{CSST Science Center for the Guangdong-Hongkong-Macau Greater Bay Area, DaXue Road 2, 519082, Zhuhai, China}

\author[0000-0002-1512-5653]{Jing Li}
\affiliation{School of Physics and Astronomy, China West Normal University, ShiDa Road 1, 637002, Nanchong, China}

\author{Yanyao Lan}
\affiliation{School of Physics and Astronomy, Sun Yat-sen University, DaXue Road 2, 519082, Zhuhai, China}
\affiliation{CSST Science Center for the Guangdong-Hongkong-Macau Greater Bay Area, DaXue Road 2, 519082, Zhuhai, China}

\correspondingauthor{Lin Tang, Weipen Lin}
\email{tanglin23@cwnu.edu.cn, linweip5@mail.sysu.edu.cn}

\label{firstpage}

\begin{abstract}

In this paper, by analyzing mock images from the IllustrisTNG100-1 simulation, we examine the properties of the diffuse light and compare them to those of central and satellite galaxies.
Our findings suggest that the majority of the diffuse light originates from satellites. 
This claim is supported by the similarity between the age and metallicity distributions of the diffuse light and those of the satellites.
Notably, the color distribution of the diffuse light gradually evolves to resemble that of the centrals at lower redshifts, suggesting a coevolution or passive process.
The radial profiles of the diffuse light reveal distinct trends, with the inner regions displaying a relatively flat distribution and the outer regions showing a descending pattern.
This finding suggests that the formation of the diffuse light is influenced by both major mergers and stellar tidal stripping.
Moreover, strong correlations are found between the stellar mass of the diffuse light and the overall stellar mass of the satellites, as well as between the stellar mass of the diffuse light and the number of satellites within groups or clusters.
These relationships can be described by power-law and logarithmic functions.
Overall, the diffuse light components predominantly originate from satellites with intermediate ages and metallicities.
These satellites typically fall within the stellar mass range of $\rm 8<\log_{10}M_{star}/M_{\odot}< 10$ and the color range of $\rm -1<[g-r]^{0.1}< 0$. 
As the redshift decreases, the growth of the diffuse light is primarily influenced by the redder satellites, while the most massive and reddest satellites have minimal roles in its growth.

\end{abstract}




\section{Introduction}\label{introduction}

In the past two decades, there has been extensive research on intracluster light (also known as diffuse light), which is a widespread and diffuse stellar component found in galaxy clusters.
Typically located at the cluster center, encompassing the brightest cluster galaxies (BCGs), the intracluster light is associated with the overall gravitational potential well of the cluster, rather than any individual galaxy.
The study of the intracluster light is crucial because it contributes significantly to our understanding of galaxy clusters and their host dark matter halos, potentially shedding light on their assembly histories {\citep[][for reviews]{Montes2019,Montes2022,Contini2021}.}

Note that the intracluster light is not only present in clusters but is also observed in groups of galaxies, as supported by simulations and observations \citep[e.g.,][]{Gonzalez2005, Gonzalez2007, Krick&Bernstein2007, Contini2019}. 
Considering that we treat both groups and clusters in this work, we use the general term ``diffuse light'' when describing the diffuse stellar component associated with the overall gravitational potential well of the groups or clusters, rather than any individual galaxy.
In contrast to the typically observed luminous galaxies, the diffuse light represents a fainter aspect of the universe. 
Exploring these faint components can provide valuable insights that enhance our understanding of galaxy formation and evolution.

The properties, formation, and evolution of the diffuse light have been extensively investigated and discussed in the literature.
For instance, the stellar mass fractions of the diffuse light exhibit a wide range across different observed filters, masses of groups and clusters, and redshifts.
\cite{Jimenez-teja2018} discovered that the fractions of the diffuse light in 11 massive clusters of galaxies, as detected by the Hubble Space Telescope, vary from $2\%$ to $23\%$ in three broadband optical filters: F435W, F606, and F814W.
Building upon the analysis of \cite{Jimenez-teja2018}, \cite{Oliveira2022} found that the diffuse light percentages for both A370 and AS1063 range from $7\%$ to $25\%$ and $3\%$ to $22\%$ of the total light of the respective clusters.
When detecting the diffuse light at different redshifts, researchers have found that the fraction of the diffuse light increases as the redshift decreases \citep[e.g.,][]{Rudick2011,Burke2015,Jimenez-teja2018, Montes&Trujillo2018,Furnell2021,Kluge2021,Montes2021} and grows alongside the assembly of groups and clusters of galaxies \citep[e.g.,][]{Mihos2016,Zhang2019,Rodrigo2020,Kluge2021,Sampaio-Santos2021}.
However, \cite{Joo&Jee2023} recently found that the diffuse light is already highly abundant in galaxy clusters with redshifts of $z \ge 1$, and  that its fraction is about twice the theoretical prediction of \cite{Rudick2011} at high redshifts.
A high fraction of the diffuse light at high redshifts is also reported in more recent work \citep{Werner2023,Jim2023}.

Furthermore, the fraction of the diffuse light is proportional to the mass of the host dark matter halos in simulations \citep[e.g.,][]{Pillepich2018a,Tang2021,Sampaio-Santos2021}, while the observations show a scattered distribution of the diffuse light fraction within halos greater than $\rm 10^{12}\ M_{\odot}$ \citep[e.g.,][]{Gonzalez2005, Gonzalez2007, Krick&Bernstein2007}. 
The results of several studies also indicate that the definition of the diffuse light significantly influences its fraction \citep[e.g.,][for reviews]{Tang2018, Contini2021, Montes2022}.

Considering that the majority of the diffuse light is located in the center of dark matter halos, we can reasonably assume a significant correlation between the evolution of the diffuse light and the BCGs \citep[e.g.,][]{Zibetti2005,Rudick2009}.
However, numerous investigations have demonstrated that the distribution of the diffuse light is broader than that of the BCGs \citep[e.g.,][]{Burke2012}.
This finding indicates a clear distinction between the diffuse light and BCGs, suggesting two primary formation mechanisms for the diffuse light: galaxy mergers and tidal stripping {\citep[e.g.,][and the references therein]{Murante2007, Purcell2007, Puchwein2010, Rudick2011, Burke2015, DeMaio2015,  DeMaio2018, Groenewald2017, Morishita2017, Montes&Trujillo2018, Contini2018, Contini2019,Contini2021,Montes&Trujillo2022}.}
These mechanisms are differentiated because one would not expect color/metallicity gradients in the BCG + diffuse light system if mergers are primarily responsible for the diffuse light formation.
Importantly, note that only a small amount of the diffuse light is formed in situ from cold gas \citep[e.g.,][]{Jordan2009,Takamiya2009,Puchwein2010}.

The question of which mechanism plays a key role in the formation of the diffuse light is crucial. 
To answer this question, researchers have conducted additional studies. 
For instance, some groups and clusters of galaxies with flat color gradients  have been found and identified \citep[e.g.,][]{Krick&Bernstein2007}.
However, distinct radial gradients of the color and metallicity of BCG + diffuse light systems have been discovered, with the BCG being redder and   richer in metals than the diffuse light \citep[e.g.,][]{Montes&Trujillo2014,Morishita2017, DeMaio2018,Montes&Trujillo2018}.
This finding suggests that stellar stripping from satellite galaxies is the primary mechanism of the diffuse light formation, as confirmed by semi-analytic models in \cite{Contini2019}.
\cite{Yoo2021} extracted the diffuse light from a fossil cluster and found that the color of the diffuse light is comparable to that of the BCG out to $\sim70 \kpc$, becoming somewhat bluer toward the outskirts.
In addition, \cite{Montes&Trujillo2022} derived the color profile of a cluster in the James Webb Space Telescope (JWST) Early Release Observations and discovered a nearly constant color profile in the inner portions and a significant dip of 0.6 magnitudes in the outer parts.
These findings can be explained by a combination of major mergers in the inner portions and tidal stripping in the outer parts.

The identification of the galaxies responsible for the generation of the diffuse light is another crucial aspect for understanding its formation.
Numerous studies have concluded that the diffuse light primarily originates from massive galaxies with stellar masses comparable to that of the Milky Way \citep[e.g.,][]{Contini2014,DeMaio2018,Contini2018}.
Further, \cite{Montes&Trujillo2018} investigated the diffuse light from six massive clusters in the Hubble Frontier Fields survey.
It is found that the average color and metallicity of the diffuse light are similar to those of the Milky Way's outskirts.  
However, \cite{Morishita2017} determined that the diffuse light more likely originates from satellites with stellar masses $\rm\log M_{star}/M_{\odot}<10$.
\cite{Contini2019} anticipated a negative radial metallicity and color gradient in the BCG + diffuse light system based on a semi-analytic model of galaxy formation.
By comparing the typical colors of the diffuse light with those of satellite galaxies, they discovered that low-mass galaxies with stellar masses in the range of $\rm9<\log M_{star}/M_{\odot}<10$ are the primary contributors to the early stages of the diffuse light formation.
In contrast, intermediate/massive galaxies with stellar masses in the range of $\rm10<\log M_{star}/M_{\odot}<11$ contribute to the growth of the diffuse light content at lower redshifts.

Only the diffuse light from a few groups and clusters of galaxies have been extensively observed, as these observations require expensive wide-field observations with high spatial resolution and very deep exposure.
Conversely, simulations can provide a large sample of groups and clusters of galaxies spanning a wide range of masses. 
As a result, simulations are a valuable tool for statistically studying the properties and formation of the diffuse light.
In this paper, we aim to investigate the relationships between the diffuse light properties and those of central galaxies (henceforth referred to as centrals), satellite galaxies (henceforth referred to as satellites), and dark matter halos, as well as to determine the formation and evolution processes of the diffuse light.
We utilize the TNG100-1 simulation of IllustrisTNG, the state-of-the-art N-body/hydrodynamical simulations, to determine which factor has the most significant influence on the diffuse light and understand the relationship between them.

The rest of this paper is organized as follows. 
In Sections \ref{simulation} and \ref{ICL_definition}, we briefly describe the simulation used and the method employed to define the diffuse light.
In Section \ref{ICL_properties}, we present the age, metallicity, and color statistics of the diffuse light, as well as the comparisons with centrals and satellites.
The radial profiles of the diffuse light properties are presented in Section \ref{profile}.
In Section \ref{ICL_sat}, we examine the relationships between the satellites and diffuse light.
Finally, we briefly discuss and further summarize our main results in Section \ref{conclusions}.

\section{Simulation}\label{simulation}
In this study, the utilized simulation is the TNG100-1 simulation of  IllustrisTNG suite \footnote{https://www.tng-project.org}.
This is a cosmological hydrodynamical simulation with $1820^3$ dark matter and $1820^3$ gas particles in a cubic box of $(110.7\Mpc)^3$.
The mass resolution of the simulation is $1.4\times10^6\Msun$ for baryon particles while the spatial resolution is set by the Plummer softening length, which is $0.74\kpc$. 
The following cosmological parameters are obtained from Planck \citep{Planck2016}: $\Omega_m=0.3089$, $\Omega_{\Lambda}=0.6911$,  $\Omega_{b}=0.0486$,  $n_s=0.9667$, $\sigma_8=0.8159$, and $h=0.6774$.
The detailed description of the database can be found in \cite{Nelson2019}. 
Refer to the paper series on TNG50/100/300 for a more detailed description of the stellar content \citep{Pillepich2018b}, galaxy clustering \citep{Springel2018}, galaxy colors \citep{Nelson2018}, chemical enrichment \citep{Naiman2018}, and magnetic fields \citep{Marinacci2018}.
Compared to the previous Illustris simulations \citep[][]{Vogelsberger2014}, the TNG100-1 simulation includes a revised active galactic nucleus (AGN) feedback model, which controls the star formation efficiency of massive galaxies \citep{Weinberger2017}, and a galactic wind model, whose feedback inhibits the efficient star formation in low- and intermediate-mass galaxies \citep{Pillepich2018b}.

To reduce the mass-resolution effect of the simulation, we select the friend-of-friend (FoF) groups with masses $\rm M_{200} > 10^{12}\Msun$ and member galaxies with stellar masses $\rm M_{star} > 10^8 \Msun$. 
$\rm M_{200}$ is the mass within the virial radius $\rm R_{200}$, which is the radius within which the average density is 200 times the critical density of the universe.
The most massive FoF group in our sample has a mass $\bm M_{200} \sim 10^{14.64}\Msun$.
To determine the evolution processes of the diffuse light, we extract FoF groups from simulation snapshots 91, 78, 67 and 50 ($z=0.1, 0.3, 0.5$, and $1.0$).
\begin{figure*}
    \centering
\includegraphics[width = 0.45\textwidth]{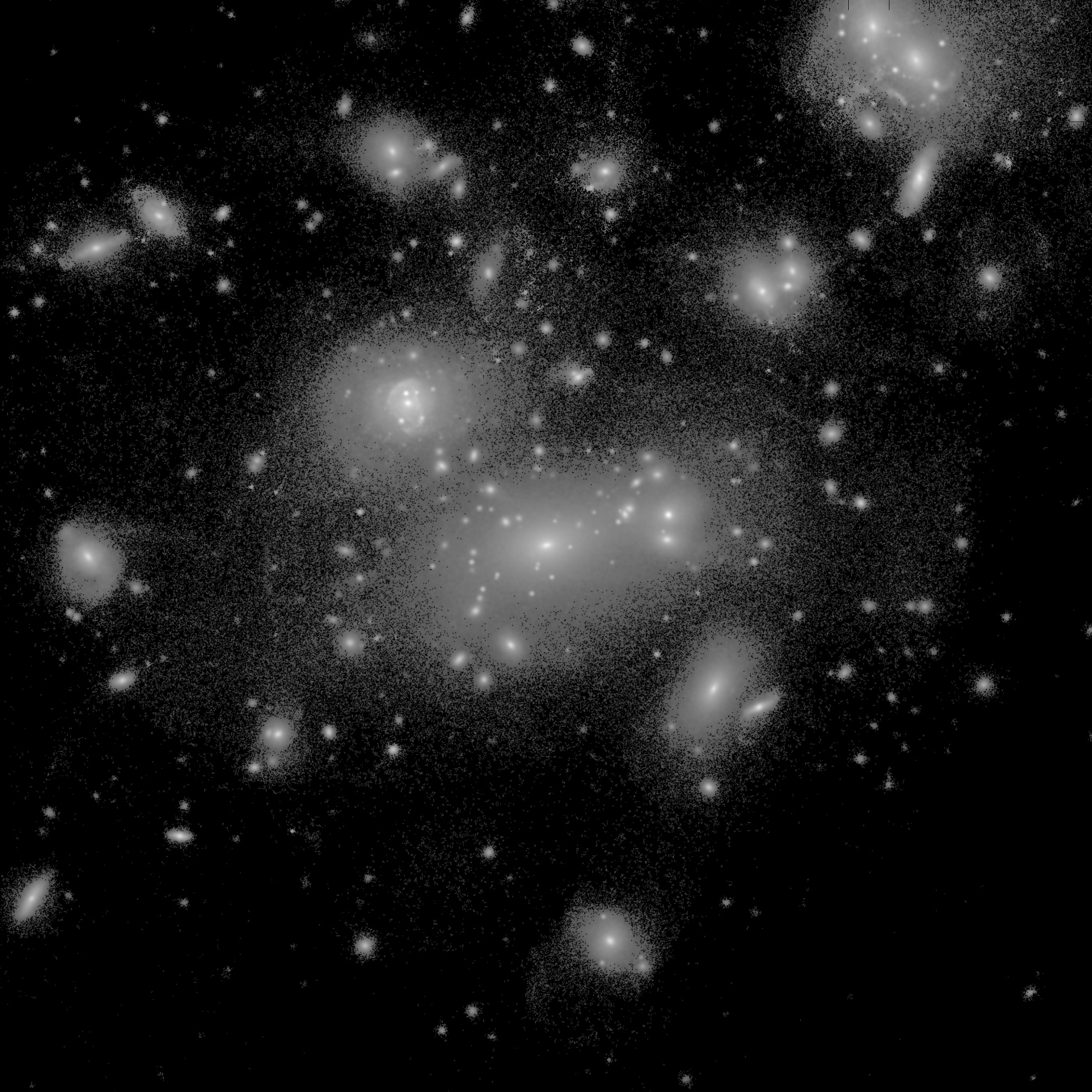}
\includegraphics[width = 0.506\textwidth]{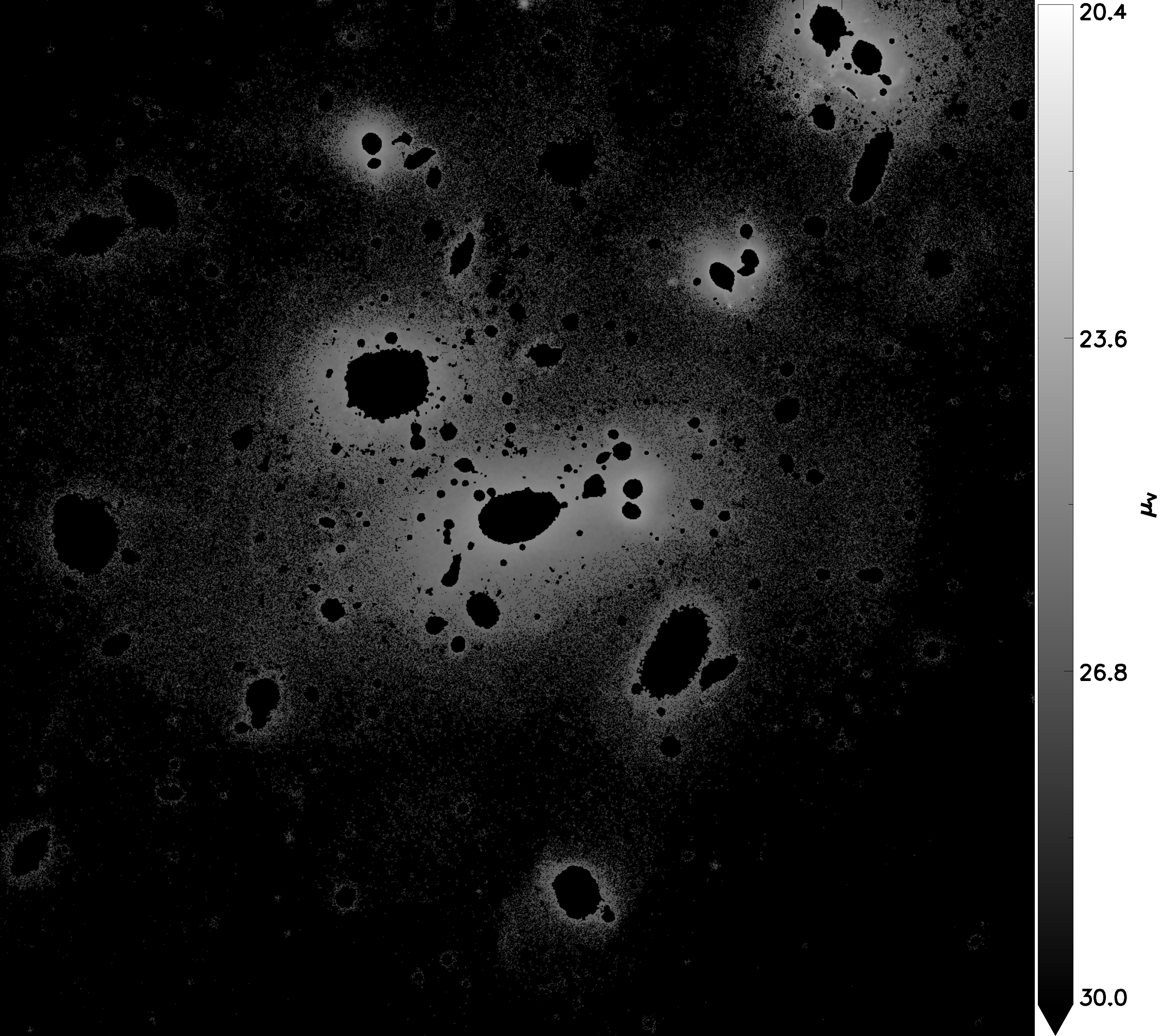}
\caption{
Illustration of luminosity profile of a dark matter halo with a virial radius of around $1.15\Mpch$ and $\rm M_{200}$ mass of $\rm2.96\times10^{14} M_{\odot}$ at redshift $z=0.1$.
The projection of this halo is shown on the x-y plane, with each side of the square being approximately $1.0\Mpch$ and centered at the center of mass.
$Left$: total luminosity profile of the halo;
$Right$: luminosity of the halo after applying a mask that excludes the galaxies defined by the SBLSP method.
The color bar represents the surface brightness in the V band.
          }
  \label{fig_illustration}
\end{figure*}
\begin{figure*}
    \centering
\includegraphics[width = 0.24\textwidth]{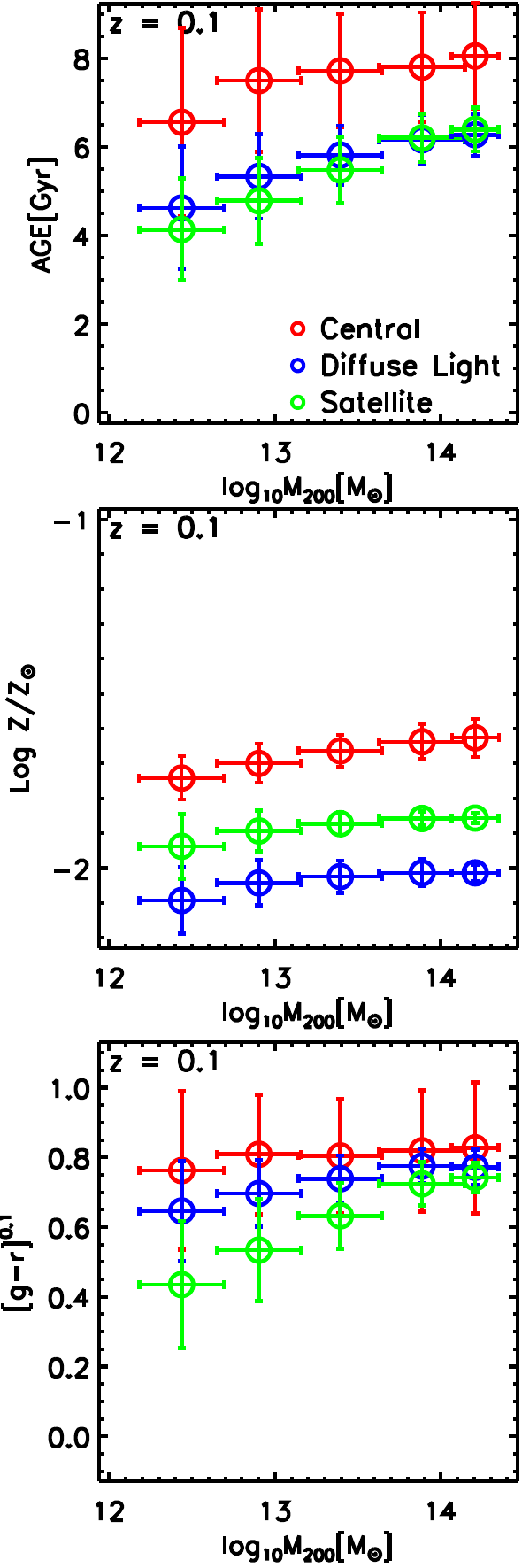}
\includegraphics[width = 0.24\textwidth]{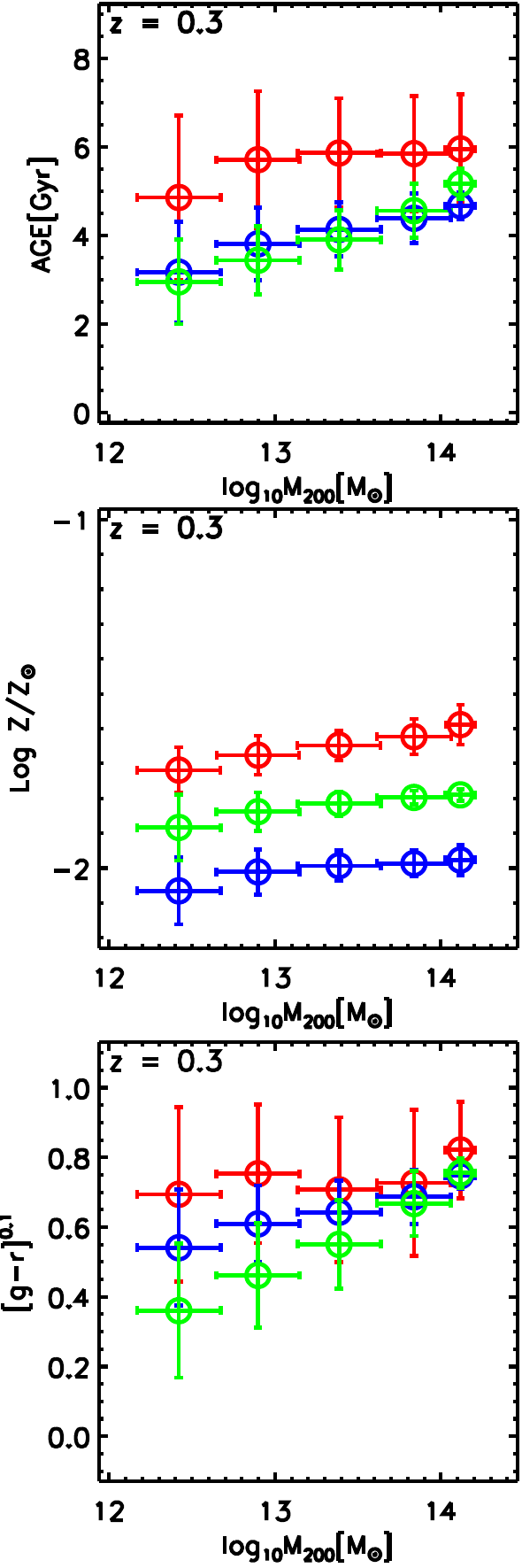}
\includegraphics[width = 0.24\textwidth]{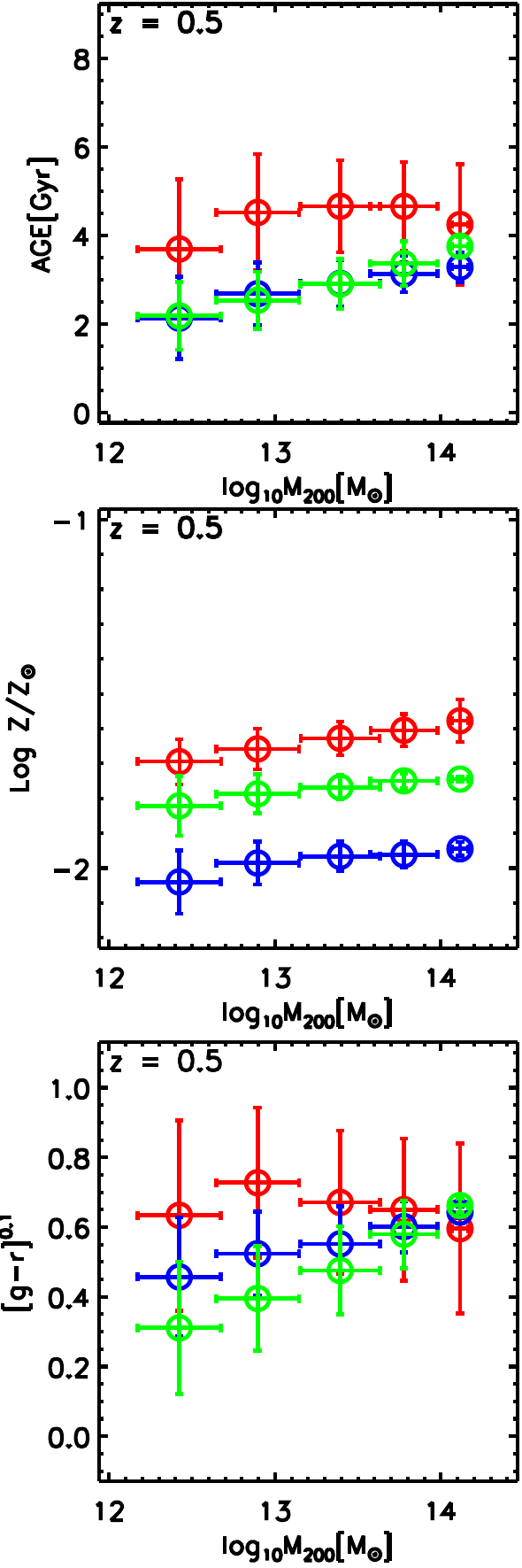}
\includegraphics[width = 0.24\textwidth]{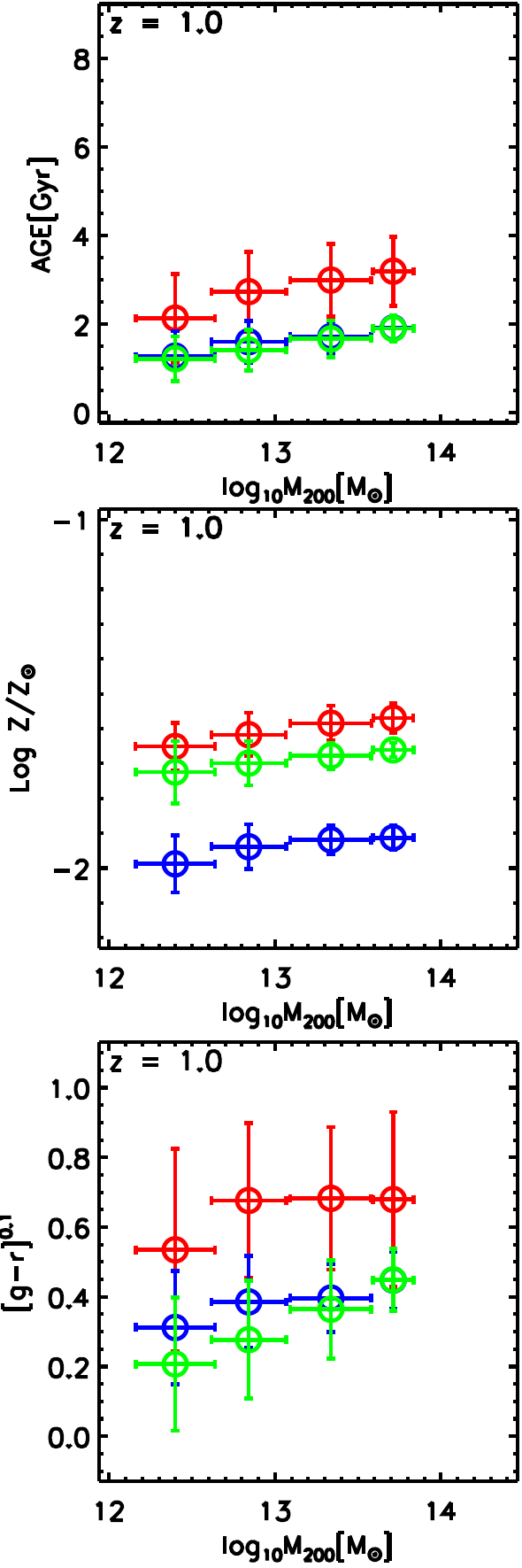}
\caption{
Relationships between intrinsic properties (age, metallicity, and color in the top, middle, and bottom panels, respectively) of the centrals (or satellites, diffuse light) and dark matter halo mass $\rm M_{200}$.
Each panel shows the results for the centrals, satellites, and diffuse light, represented by the red, green, and blue open circles, respectively.
The panels are arranged from left to right to show the results at redshifts  $z=0.1, 0.3, 0.5$, and $1.0$. 
The error bars represent the standard deviations.
            }
  \label{fig_properties}
\end{figure*}
\begin{figure*}
    \centering
\includegraphics[width = 0.24\textwidth]{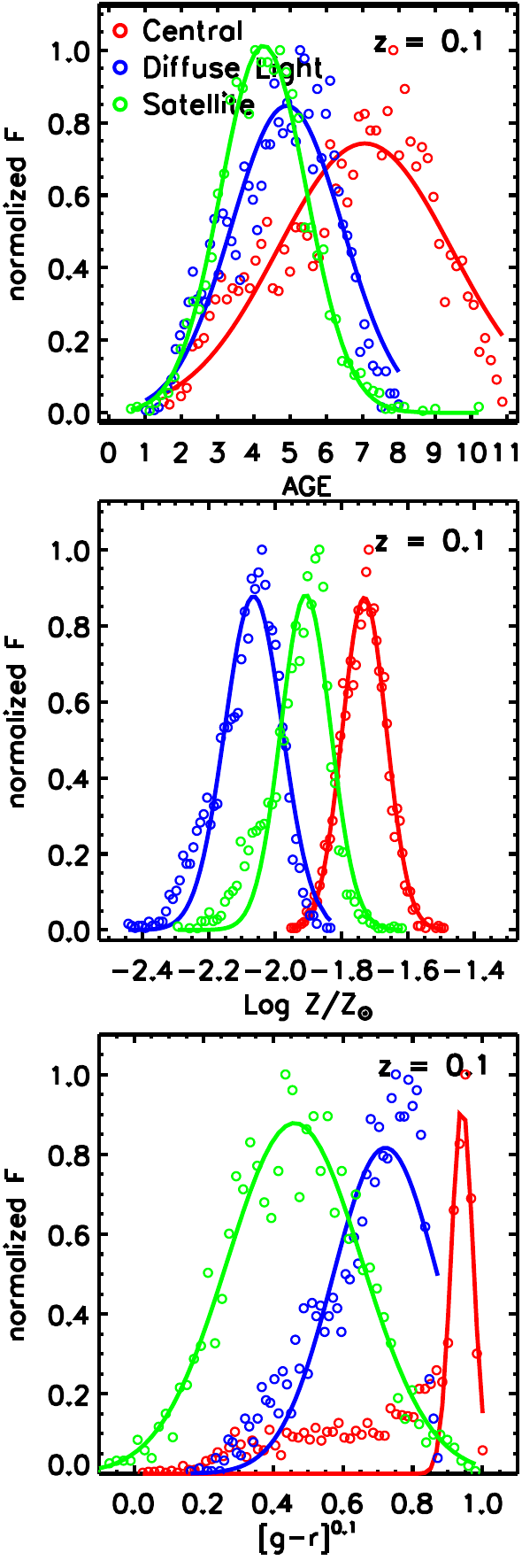}
\includegraphics[width = 0.24\textwidth]{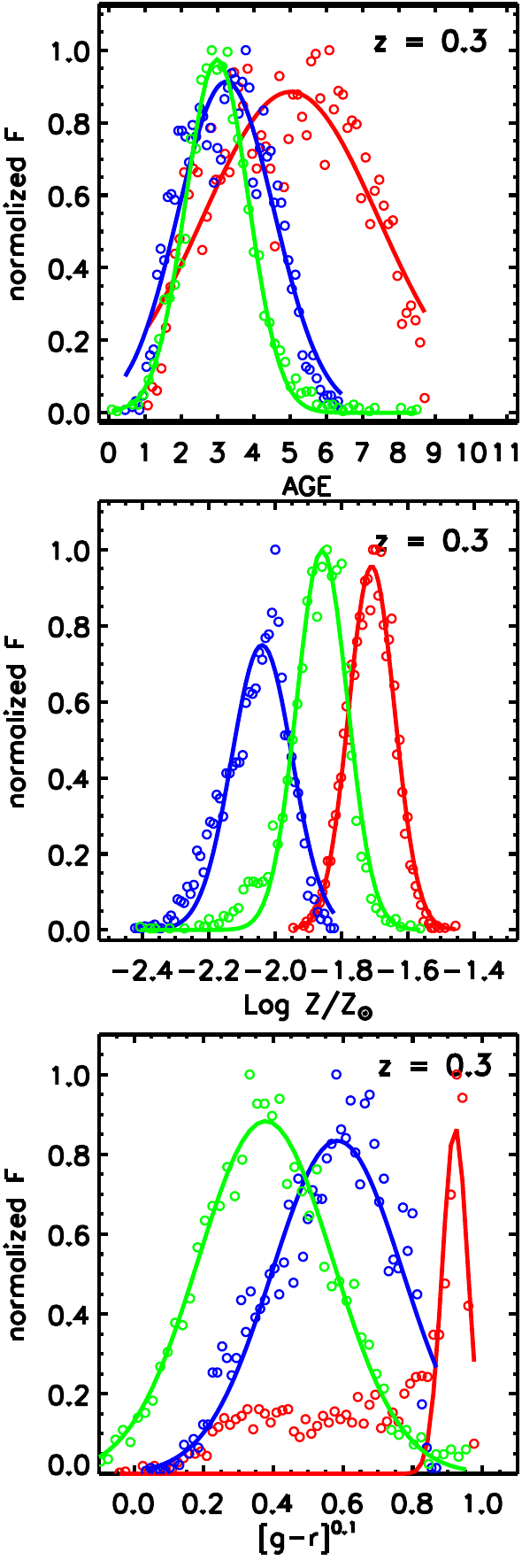}
\includegraphics[width = 0.24\textwidth]{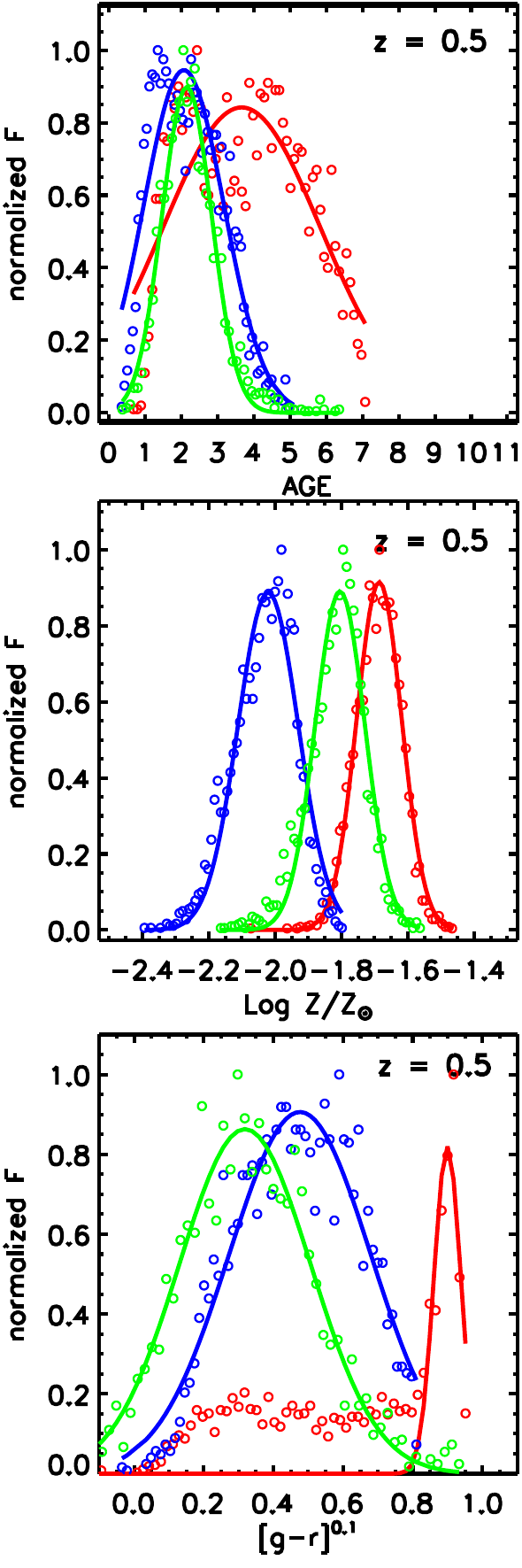}
\includegraphics[width = 0.24\textwidth]{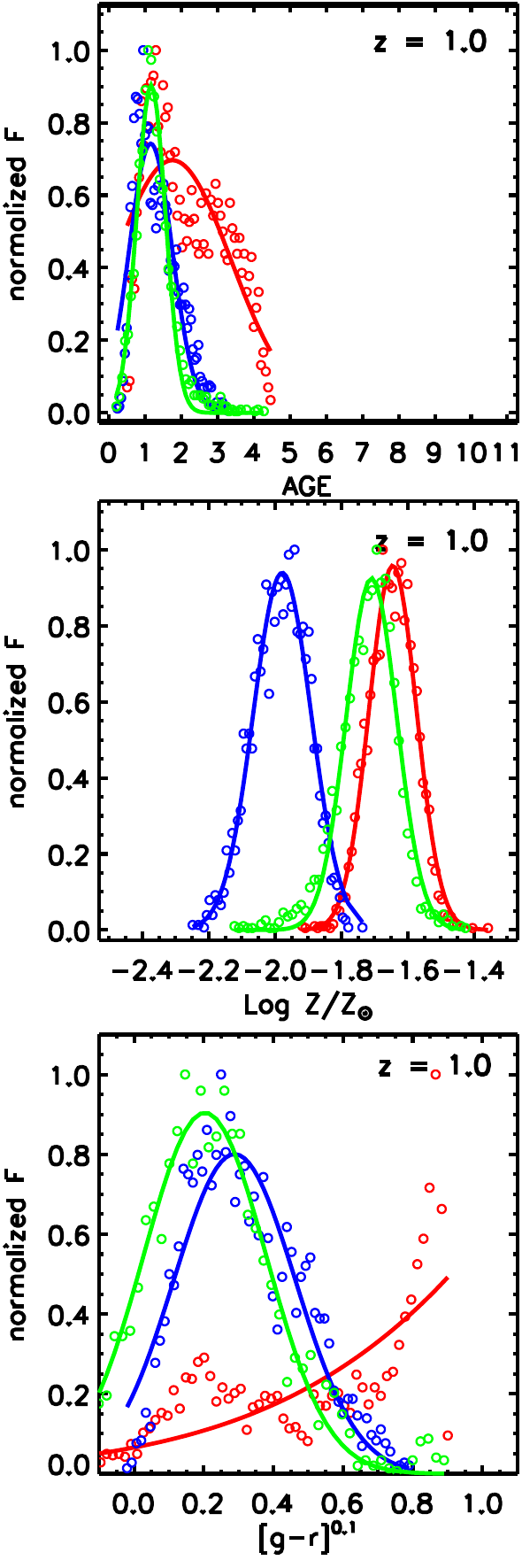}
\caption{
Number distributions of galaxy properties are shown in each panel, with the fraction normalized by dividing the maximum number in the value bins. 
The solid lines represent Gaussian fits to the distributions. 
In each panel, the red, green, and blue open circles represent the results for the centrals, satellites, and diffuse light, respectively. 
The panels from left to right correspond to the results at redshifts $z = 0.1$, $0.3$, $0.5$, and $1.0$, respectively.
            }
  \label{fig_distribution}
\end{figure*}

\section{Diffuse light definition}\label{ICL_definition}

Multiple definitions of the diffuse light exist \citep[][for review]{Arnaboldi2004, Tang2018, Montes2022}. 
In observational studies, the diffuse light is commonly defined as the excess of light beyond the profile of a BCG when removing the sky background and satellite galaxies \citep[e.g.,][]{Gonzalez2000,Gonzalez2005, Zibetti2005, Seigar2007,Jee2010, Melnick2012, Giallongo2014,Spavone2018,Spavone2020}. 
However, the choices of the fitting method and outer fitting radius can lead to variations in the results.
Another approach is the surface brightness (SB) threshold method, where the stellar component with a surface brightness fainter than a certain threshold, typically the observational surface brightness limit, is defined as the diffuse light \citep[e.g.,][]{Zibetti2005, Puchwein2010, Presotto2014}.

In \cite{Tang2018}, the evolution of the observational light fraction of the diffuse light was reproduced by the SB threshold method in simulations.
In \cite{Tang2021}, it was found that the stellar mass in the mock galaxy sample resembles the observations more closely than approaches based on sub-finder structures \citep{Springel2001} and aperture-selected galaxies \citep{Pillepich2018a}.
To obtain more accurate results, we define the diffuse light based on mock galaxy images \citep{Tang2018}, effectively eliminating the influences of the components of the central and field galaxies extracted by the surface brightness level segmentation procedure (SBLSP) introduced in \cite{Tang2020,Tang2021}, which is based on SB thresholds to distinguish between the member galaxies and the diffuse light in the TNG100-1 simulation of IllustrisTNG.

Initially, we generate mock images of the simulated FoF groups on three projecting planes ($xy$, $yz$, $yz$) by considering various factors, such as the point-spread function (PSF) characterized by a Moffat function \citep{Moffat1969, Trujillo2001}, the pixel size (D), surface brightness limit (SBL) of the image, spatial resolution, dust extinction, and redshift dimming effects.
During this process, the following astrophysical quantities can be calculated for each projecting mesh: (1) the total stellar mass, obtained by summing the masses of all star particles within each mesh; (2) the total luminosity, obtained by summing the luminosities of all the star particles in four filters ($B$, $V$, SDSS $g$, and SDSS $r$) within each mesh; (3) $\rm[g-r]^{0.1}$ color, calculated by applying the transformation $\rm[g-r]^{0.1}=0.7188+1.3197\lbrack(g-r)-0.6102\rbrack$ \cite[]{Blanton&Roweis2007} to correct the color at $z=0.1$; (4) the mean age and metallicity, calculated with weights based on the stellar luminosity; and (5) the sum of the individual star formation rates of all gas cells within the mesh. 
Note that there is no age of the stellar particles in the IllustrisTNG datasets, which can be defined as the cosmic time at the snapshot redshift  minus the cosmic time of the stellar particle formation listed in the IllustrisTNG datasets. 
The cosmic time is calculated using the functions \citep{Mo2010},
\begin{equation}
	\label{eq_cosmostime1}
	t(z)=\int_{0}^{a(z)}\frac{{\rm d}a}{\dot{a}}=\frac{1}{H_{0}}\int_{z}^{\infty}\frac{{\rm d}z}{(1+z)E(z)},
\end{equation}
\begin{equation}
	\label{eq_cosmostime2}
       \begin{split}
	E(z)=[\Omega_{\Lambda,0}+(1-\Omega_{0})(1+z)^2+\\
	\Omega_{m,0}(1+z)^3+\Omega_{r,0}(1+z)^4]^{1/2},
	 \end{split}
\end{equation}
where $a$ is the scale factor and $z$ is the redshift, $\Omega_{m,0}=0.3089$, $\Omega_{\Lambda,0}=0.6911$,  $\Omega_{r,0}=0$,  $\Omega_{0}=1$, and $H_{0}=100h$.
The stellar age is defined as $t(z_{\rm snap})-t(z_{\rm sft})$, where $t(z_{\rm snap})$ and $t(z_{\rm sft})$ are the cosmic time of star formation and the snapshot, respectively.

The surface brightness for each pixel is given by
\begin{equation}
	\label{eq_ux}
	\mu_x=-2.5\log\frac{I_x}{L_{\odot,x} \cdot pc^{-2}}+21.572+M_{\odot,x},
\end{equation}
where $I_{x}$ is
\begin{equation}
	\label{eq_lx}
	I_x=\frac{L_x}{\pi^2 D^2}(1+z)^{-4}.
\end{equation}
Here, $x$ indicates different filters in the rest frame, and $\rm M_{\odot,x}$ is the absolute solar magnitude, with values of $5.36, \ 4.80, \ 5.12,$ and $\ 4.64\ \rm mag/arcsec^{-2}$ \citep{Blanton&Roweis2007} for the $B$, $V$, SDSS $g$, and SDSS $r$ filters in the AB system, respectively.

Next, we separate the diffuse light components from the member galaxies by applying a given SB threshold.
To clearly delineate the boundaries of the galaxies with different SB levels, as shown in Figure 6 of \cite{Tang2020}, we repeat this step with a series of SB thresholds to obtain sample templates.
Subsequently, we utilize a segmentation procedure, the SBLSP, which is introduced in detail in \cite{Tang2020, Tang2021}, to obtain the final galaxy sample from the sample templates defined by various SB thresholds.
Finally, the diffuse light refers to the stellar particles that do not belong to galaxies.
All the galaxies in our samples are chosen with stellar masses greater than $10^{8}\Msun$.

Note that the softening length is set as the fiducial `simulation-CCD' pixel size of the projected images.
For the charge-coupled diode (CCD) pixel, the variance of the Poisson distribution is set to 30 with units of $\rm mag/arcsec^{-2}$ and the magnitude of the CCD pixel, representing the sky background and CCD noise, respectively.
Additionally, we convert the softening length into an angular resolution to serve as a `simulation-PSF'.
In the SBLSP method, we employ the radiative transfer code SKIRT \citep[][]{Baes2011} to add a dust component to the star-forming mock galaxies. 
The observational limit is set at 26.5 $\rm mag/arcsec^{-2}$.
The series of SB thresholds used in the second procedure range from 18 to 26.5 $\rm mag/arcsec^{-2}$, with surface-brightness intervals of 0.1 $\rm mag/arcsec^{-2}$.

In Figure \ref{fig_illustration}, we present the diffuse light within a dark matter halo at redshift $z=0.1$, as defined by the SBLSP method.
The halo has an $\bm R_{200}$ of $\rm 1.15\Mpc$ and an $\rm M_{200}$ mass of $2.96\times10^{14}\Msun$. 
The x-y projected plane has a central region of $1.0\ \Mpch \times1.0\ \Mpch$.
The left panel of Figure \ref{fig_illustration} illustrates the total luminosity profile of the halo, while the right panel shows the luminosity profile after applying a mask to exclude member galaxies.
All the peaks in the total luminosity profile presented in the left panel are eliminated due to the masking of member galaxies defined by the SBLSP method. 
Consequently, the remaining luminosity is significantly faint and primarily exhibits a surface brightness in the V band of $\rm\mu_{V}\lesssim25 \ mag/arcsec^{-2}$.
As shown in the right panel of Figure \ref{fig_illustration}, the majority of the diffuse light is concentrated in the center of the dark matter halo. 
Additionally, a large satellite object is situated close to the halo center, surrounded by the diffuse light. 
In the top-right corner of the right panel, a considerable amount of the diffuse light is found, which is believed to represent remnants from a falling halo. 
Notably, dark stellar streams can be identified between the centers of dark matter halos and neighboring member galaxies.


\section{Diffuse light properties and comparison with centrals and satellites}\label{ICL_properties}
In this section, we present a statistical analysis of the diffuse light properties, including the age, metallicity, color, and stellar mass. 
We also examine the centrals and satellites for comparison.
Those properties of the three components are calculated in the same way for each mesh, as described in Section \ref{ICL_definition}.
Note that we define the most massive galaxies located in the dark matter halos as the central galaxies and the remaining galaxies as the satellite galaxies.

\begin{figure*}
\centering
\includegraphics[height=5.8cm,width=18cm]{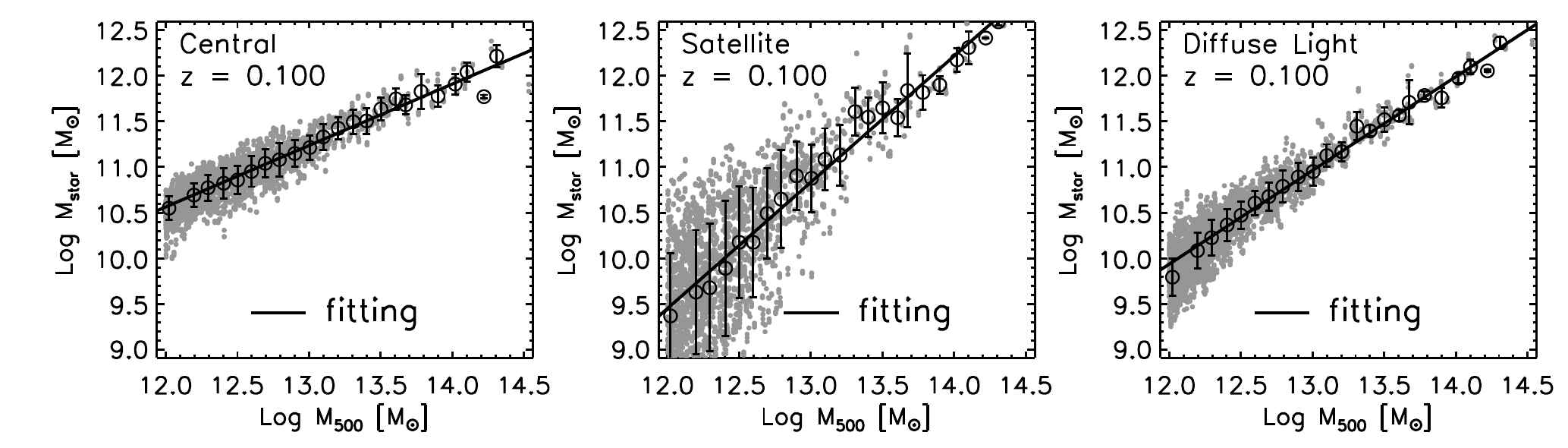}
\caption{
Relationships between the dark halo mass $M_{500}$ and the stellar masses of the centrals (left panel), total satellites (middle panel), and diffuse light (right panel) within dark matter halos at redshift $z=0.1$.
The gray points represent the results for each component in the dark matter halos.
The open circles with error bars indicate the average values within the mass bins, and the solid lines represent the fitting results.
The error bars show the standard deviations of the data points.
            }
\label{fig_MR}
\end{figure*}

Figure \ref{fig_properties} illustrates the relationships between the dark matter halo mass $\rm M_{200}$ and the age, metallicity, and color for the diffuse light, centrals, and satellites.
Figure \ref{fig_distribution} displays the normalized number distributions of these intrinsic properties (age, metallicity, and color).
To demonstrate the redshift evolution, we present the results of four redshifts ($z=0.1$, $0.3$, $0.5$, and $1.0$) from left to right in Figure \ref{fig_properties} and \ref{fig_distribution}.
Additionally, Figure \ref{fig_MR} shows the relationships between the dark matter halo mass and the stellar masses of the diffuse light, centrals, and satellites.
{Due} to their similarity, only the results at $z = 0.1$ are shown in Figure \ref{fig_MR}.
\subsection{Age, metallicity, and color}
The top panels of Figure \ref{fig_properties} reveal that the age of the centrals remains nearly constant as the mass of the dark matter halos  increases, but slightly increases at redshift $z=1.0$.
However, the older diffuse light and satellites tend to be found in more massive halos at lower redshifts.
In the middle panels of Figure \ref{fig_properties}, a positive correlation is found between the metallicities of the centrals, satellites, and diffuse light and the masses of the dark matter halos.
In the bottom panels of the same figure, it can be found that the color of the centrals remains fairly constant, while the redder diffuse light and satellites are more prevalent in the more massive dark matter halos.

From the top panels of Figure \ref{fig_properties}, note that on average, the diffuse light components are younger than those of the centrals and close to those of the satellites.
This trend continues from redshift 1 to 0.1.
This distinction between the age of the diffuse light and centrals is further highlighted in the top panel of Figure \ref{fig_distribution}, where the ages of the diffuse light resemble those of the satellites more than those of the centrals.
As shown in the middle panels of Figure \ref{fig_properties}, the diffuse light components exhibit significantly lower metallicity than the satellites and centrals.
Also, note that while the metallicities of the diffuse light and centrals show little evolution, satellites exhibit a stronger trend of decreasing average metallicity, with their metallicities falling between those of the diffuse light and centrals, as shown in the middle panels of Figures \ref{fig_properties} and \ref{fig_distribution}.
This finding can be explained by the fact that early infalling satellites, which are similar in metallicity to the centrals during the major merger stage, have either fully merged or their remaining cores are too faint to be detected. 
However, late infalling satellites represent younger, metal-poor galaxies.
This finding is consistent with those of previous studies \citep[e.g.,][]{Engler2018}, which also indicates that early infalling satellites are relatively more metal rich than late infalling satellites.

The bottom panels in Figures \ref{fig_properties} and \ref{fig_distribution} reveal a different trend than those in the top and middle panels.
In contrast to the age and metallicity results, where satellites fall between the diffuse light and centrals, the color of the diffuse light falls between the satellites and centrals, while all three components become redder with decreasing redshift.
Importantly, the color distribution of the diffuse light is closer to that of the satellites at higher redshift and becomes closer to that of the centrals at lower redshift , which can be found in the bottom panels of Figure \ref{fig_distribution}.

The results of the color distribution can be explained as follows.  
At redshifts $z>1$, the stars in the diffuse light components are newly formed \citep[e.g.,][]{Burke2012,Ko&Jee2018,Joo&Jee2023}, either stripped from satellites or formed in situ, resulting in a bluer color compared to that of the satellites. 
However, as the diffuse light components evolve toward lower redshifts ($z\le 1$), the formation of the diffuse light decreases significantly, with only a few stars still forming (possibly in situ), while the majority of stars in the diffuse light are dying.
As a result, the diffuse light rapidly undergoes quenching and shifts toward a red color due to the strong physical mechanisms, in the densest regions of the dark matter halos. 
In comparison, even though satellites are quenched, the presence of ongoing star formations leads to a relatively blue color.
Also, note that the color distribution for the diffuse light and satellite appears broad in Figure \ref{fig_distribution}, whereas the majority of the centrals have very red colors due to significant quenching caused by the AGN and stellar feedback, which is determined by the simulation settings.
However, the long tail in the color distribution of centrals, along with the presence of a second peak or tail in the age distribution, indicates that a portion of the centrals still exhibit a blue color, suggesting the ongoing formation of young stars from the cold gas.
\begin{figure*}
\centering
\includegraphics[height=5.8cm,width=18cm]{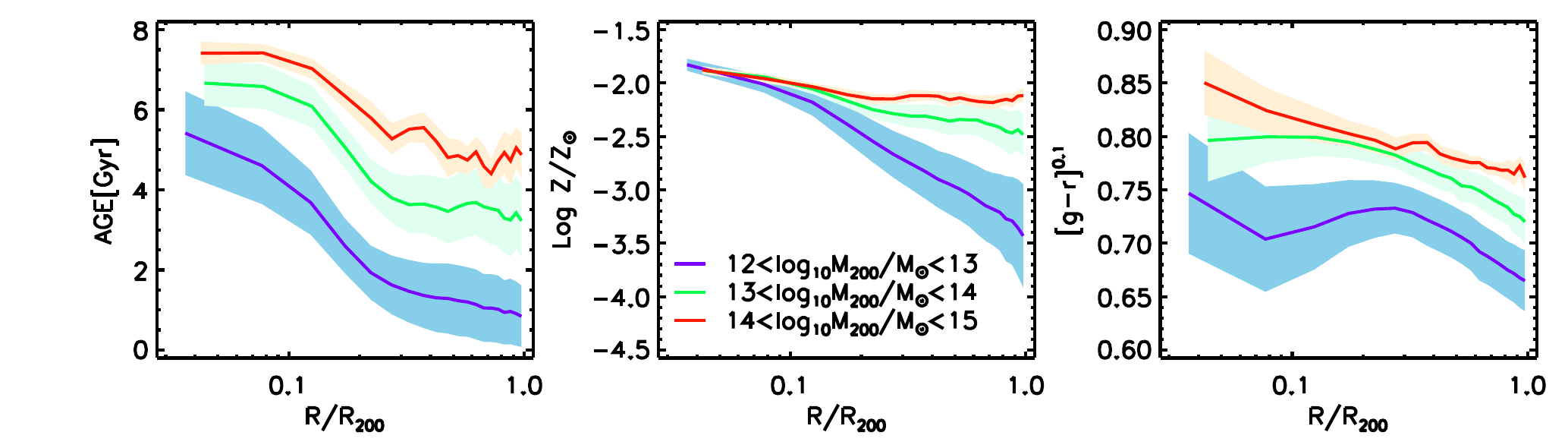}
\caption{ 
Radial profiles of intrinsic properties for the diffuse light, with the age, metallicity, and color shown from left to right.
The lines in different colors (blue, green, and red) represent the profiles within halos of different mass ranges ($12<{\rm log_{10}M_{200}/\Msun}<13, 13<{\rm log_{10}M_{200}/\Msun}<14, 14<{\rm log_{10}M_{200}/\Msun}<15$).
The error---indicated by sky blue, light cyan, and papaya---represent the standard deviations.
            }
\label{fig_radial}
\end{figure*}
\begin{table}
\centering
\begin{tabular}{cc|cc|cc|cc}
\hline
\multicolumn{2}{c|}{\multirow{2}{*}{Redshift}} & \multicolumn{2}{c|}{{Central}--Halo}&\multicolumn{2}{c|}{Satellite--Halo}&\multicolumn{2}{c}{Diffuse light--Halo}\\
\cline{3-8}
\multicolumn{2}{c|}{}     & a & b& a&b  & a & b \\
\hline
\multicolumn{2}{c|}{Combination$^A$} & 0.51 & 11.52 & 1.30 & 12.21 & 1.08 & 12.17 \\ 
\hline
\multicolumn{2}{c|}{0.1} & 0.67 & 11.91 & 1.38 & 12.21 & 1.02 & 11.99 \\ 
\hline
\multicolumn{2}{c|}{0.3} & 0.66 & 11.82 & 1.34 & 12.18  & 1.00 & 12.07 \\ 
\hline
\multicolumn{2}{c|}{0.5} & 0.69 & 11.80 & 1.35 & 12.16  & 1.00 & 12.13 \\ 
\hline
\multicolumn{2}{c|}{1.0} & 0.70 & 11.71 & 1.33 & 12.10 & 1.00 &12.26 \\ 
\hline
\end{tabular}
\caption{
Best-fitting parameters of stellar mass--halo mass relationships for the centrals, satellites, and diffuse light.
\\
A: combination results at redshifts $z=0.01, 0.1, 0.2$ in \cite{Tang2021}.
}
\label{fitting_para}
\end{table}

In conclusion, our analysis reveals that the diffuse light components differ from those of the centrals but resemble those of the satellites in terms of their relatively young ages, poor metallicities, and red colors.
This finding suggests that the diffuse light components are primarily formed through merger events or tidal stripping of infalling satellites but  evolve passively with the centrals.
Furthermore, the ages, metallicities and colors for the diffuse light, satellites, and centrals show positive correlations with the halo mass, with the diffuse light components displaying the highest sensitivity.
This finding indicates that more massive halos tend to host older, metal-richer and redder centrals, satellites, and diffuse light.
Note that the dips in the most massive bins are likely a result of limited statistical data.


\subsection{Stellar mass}
\cite{Tang2021} investigated the relationships between the dark matter halo mass ($\rm M_{500}$) and the stellar mass of the diffuse light, satellites, and centrals, and found that the results are consistent with the observational findings.
$\rm M_{500}$ here denotes the mass within the virial radius $\rm R_{500}$, which is the radius within which the average density is 500 times the critical density of the universe.
As reported in \cite{Tang2021}, the correlation between the dark matter halo mass and stellar mass is weaker for the centrals than for the satellites.
The fitting parameters for the diffuse light and satellites show similar results, with slopes of 1.08 and 1.30 and intercepts of 12.17 and 12.21, respectively.
However, the fitting parameters for the centrals are notably different, with a slope of approximately 0.5 and an intercept lower than 12.00.
Previous results show significant scatter due to the combination of data from three redshifts ($z=0.01, 0.1,$ and $0.2$).
In this work, we specifically examine the stellar mass--dark matter halo mass relationships for a set of redshifts---namely $z = 0.1, 0.3, 0.5$, and $1.0$---to investigate the redshift evolution.

The results are presented in Figure \ref{fig_MR}, and the parameters of best-fit are listed in Table \ref{fitting_para}.
Due to the similarity, only the results at $z = 0.1$ are displayed in Figure \ref{fig_MR} for viewing.
The fitting equation is the same as that in \citet{Pillepich2018a}, which is given as follows,
\begin{equation}
	\label{eq_fit_MR}
	y = a m + b,
\end{equation}
where $m={\rm log}_{10}( \MF/ \Msun) -14$, $y = {\rm log}_{10}(\rm M_{\rm stars} / \Msun)$, $a$ is the slope and $b$ is the intercept.

The results obtained from our analysis demonstrate that the stellar mass--dark matter halo mass relationships for the centrals, satellites, and diffuse light exhibit strong correlations, and that these relationships remain relatively unchanged across the examined redshifts.
The slopes $a$ and intercepts $b$ for the central stellar mass--dark matter halo mass relationship are consistently around 0.6 and 11.8 at all four redshifts. 
In comparison, the satellites and diffuse light have larger slopes and intercepts than the centrals, despite having similar fitting parameters.
The satellites have a slope of approximately 1.3, while the diffuse light has a slope of approximately 1.0. 
Both the satellites and diffuse light have intercepts of approximately 12.1. 
These findings are consistent with the results presented in \cite{Tang2021}.
\begin{figure*}
\centering
\includegraphics[width = 0.85\textwidth]{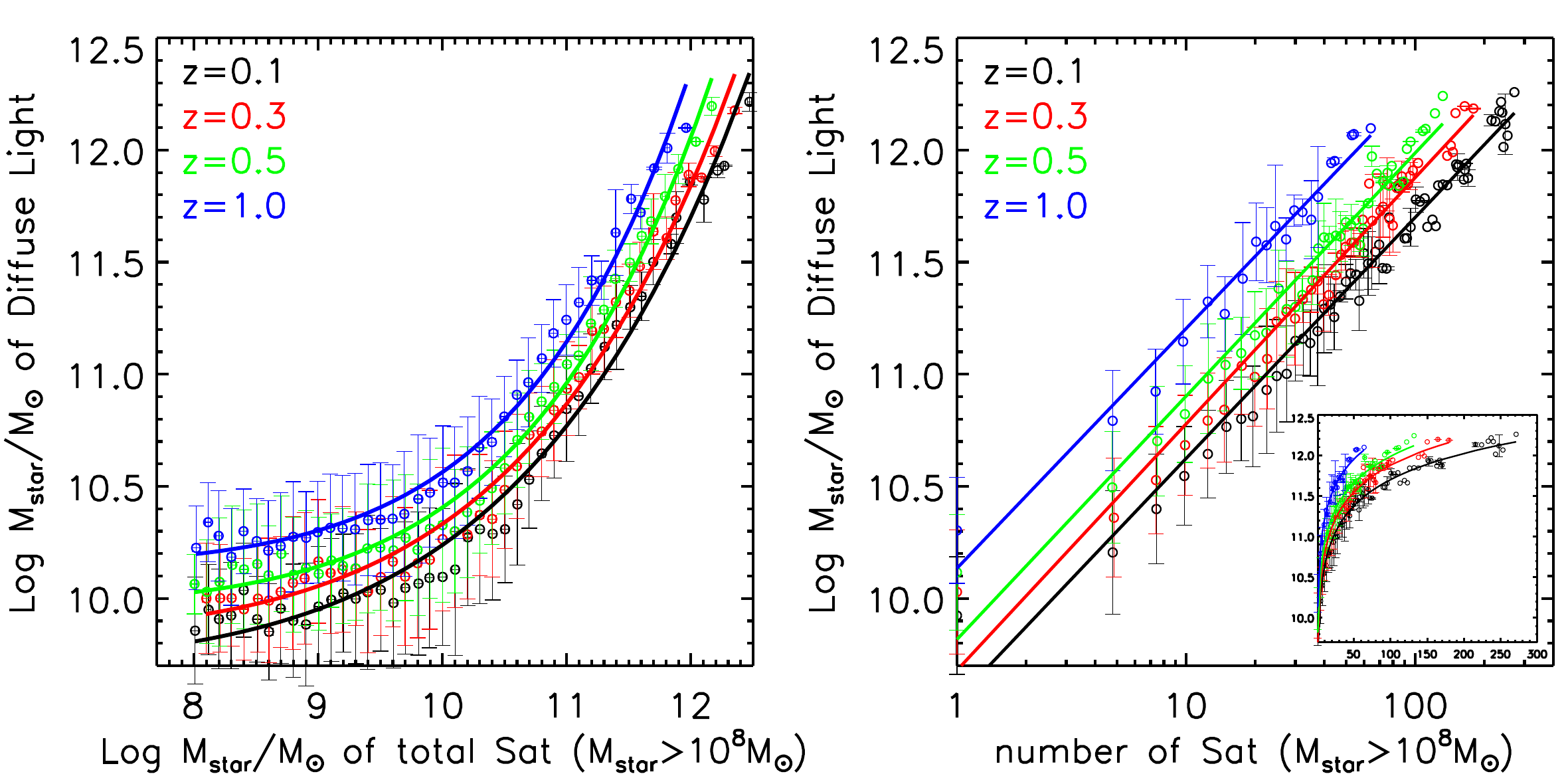}
\caption{ 
$Left$: relationship between the stellar mass of the diffuse light and the total stellar mass of satellites that are more massive than $10^{8}\Msun$.
$Right$: relationship between the stellar mass of the diffuse light and the number of satellites that are more massive than $10^{8}\Msun$.
These relationships are analyzed at different redshifts---$z = 0.1, 0.3, 0.5$, and $1.0$---represented by the colors black, red, green, and blue, respectively.
Our results are depicted by the open circles with error bars, where the error bars represent the standard deviation within each mass bin.
The solid lines represent fitting lines using different functions.
The fitting functions and parameters can be found in Table \ref{fit_Sat_number_R}.
Note that the inset in the right panel provides the same results but with a linear axis for the satellite number, instead of a logarithmic scale.
            }
\label{fig_Sat_number_R}
\end{figure*}
\begin{figure}
\centering
\includegraphics[width = 0.4\textwidth]{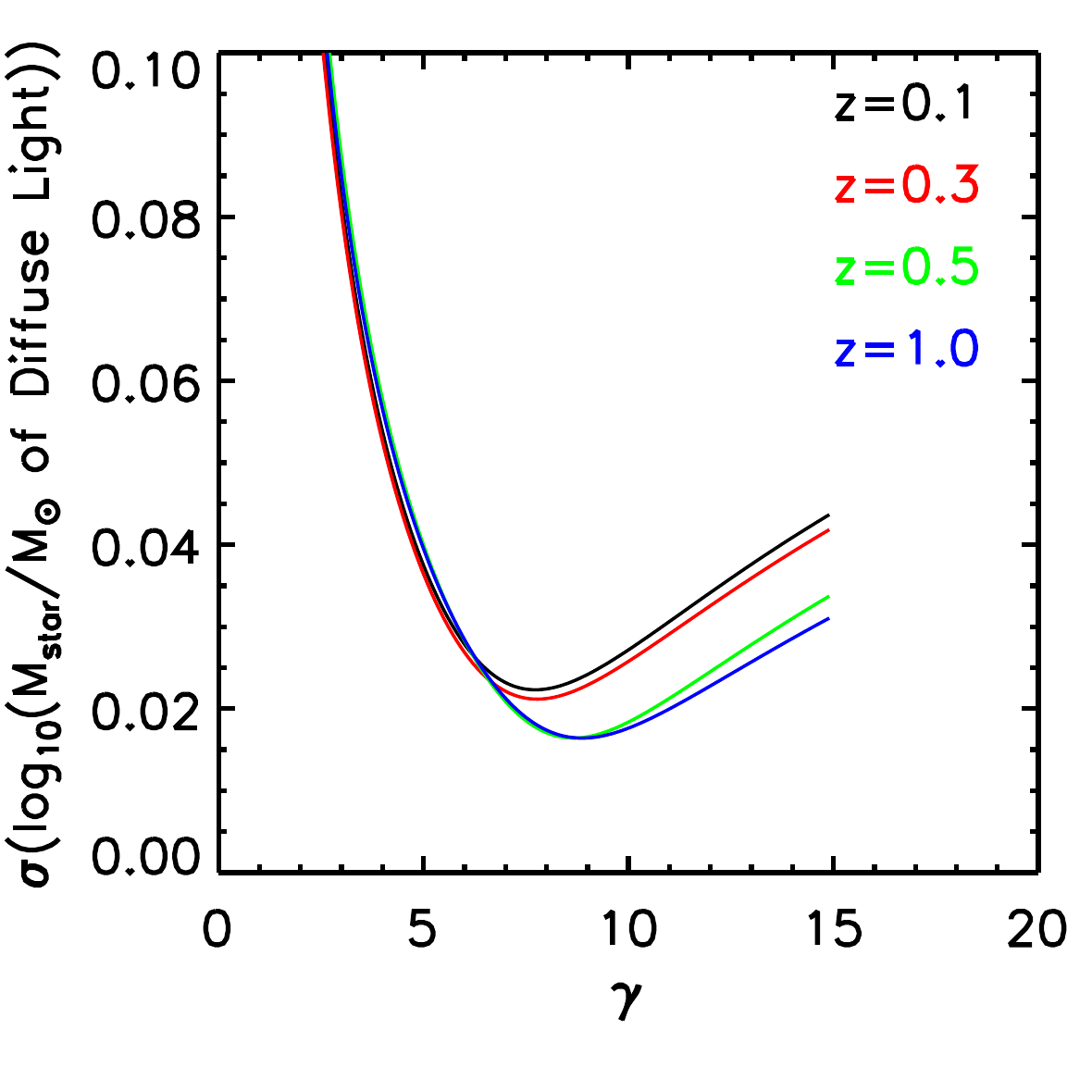}
\caption{ 
Dispersion of the best-fitting Equation \ref{eq4} in left panel of Figure \ref{fig_Sat_number_R} as a function of $\gamma$.
Specifically, we focus on dispersions in $\log_{10}({\rm M_{star}\ of\ diffuse \ light})$, as the dispersions in $\rm \log_{10} M_{star} \ of \ satellites$ are nearly 0 for $\gamma > 2$.
            }
\label{sigma_check}
\end{figure}

\section{Radial profiles of diffuse light}\label{profile}
Several studies have reported an apparent radial trend in the metallicities and colors of the diffuse light, showing an outward tilt \citep[e.g.,][]{Montes&Trujillo2014,Montes&Trujillo2018,Morishita2017,Contini2019,Edwards2020}. 
However, \cite{Yoo2021} found a similarity in the color between the diffuse light and BCG out to a radius of $70\kpc$ for a massive fossil cluster at an intermediate redshift based on deep imaging with Gemini/GMOS-N.
Furthermore, \cite{Montes&Trujillo2022} conducted an analysis of the BCG + diffuse light color profile using the JWST Early Release Observations, and found an almost constant color profile within the inner approximately $100 \kpc$.

As presented in Figure \ref{fig_properties}, our previous analysis in Section \ref{fig_properties} reveals that the centrals exhibit a redder, older, and  more metal-rich nature than the diffuse light.
Consequently, when considering both locations, we find that the centrals + diffuse light property profiles show a decline toward the outer regions.
In this section, we aim to validate the observed constant profile in the inner region by examining the radial profiles of the intrinsic properties of the diffuse light in the simulations.
Figure \ref{fig_properties} also indicates that a slight correlation exists between the properties of the diffuse light and dark matter halo mass.
To further investigate this, we divide the dark matter halos into three subsamples based on the $\rm M_{200}$ values of the dark matter halos: $12<{\rm log_{10}M_{200}/\Msun}<13, 13<{\rm log_{10}M_{200}/\Msun}<14$, and $14<{\rm log_{10}M_{200}/\Msun}<15$.
For the radial profiles of the diffuse light, we calculate the average age, metallicity, and color within redial bins from $R$ to $R+\Delta R$, where $\Delta R$ is equal to $\bm0.05R_{200}$.
In Figure ~\ref{fig_radial}, we present the radial profiles of the diffuse light properties, specifically at redshift $z=0.1$.
Note that the radial profiles at higher redshifts exhibit similar trends to those at $z = 0.1$, but they are explicitly discussed in the text.

\begin{table}
\centering
\begin{tabular}{c|cccc|cc}
\hline
\multicolumn{2}{c|}{$z$} & $\alpha_1$ & $\beta_1$ & $\gamma$ & $\alpha_2$ & $\beta_2$  \\
\hline
\multicolumn{2}{c|}{0.1} & 0.54  & 9.70 & 7.2 & 1.08 & 9.55 \\ 
\hline
\multicolumn{2}{c|}{0.3} & 0.50 & 9.83 & 7.6 & 1.10 & 9.68\\ 
\hline
\multicolumn{2}{c|}{0.5} & 0.45 & 9.96 & 8.5 & 1.08 & 9.82  \\ 
\hline
\multicolumn{2}{c|}{1.0} & 0.42  & 10.14 & 9.1 &1.07 & 10.13 \\ 
\hline
\end{tabular}
\caption{
Best-fitting parameters in Figure \ref{fig_Sat_number_R}. 
{$\alpha_1$, $\beta_1$, and $\gamma$ correspond to the left panel, while $\alpha_2$ and $\beta_2$ correspond to the right panel.
Fits are in the form of Equation \ref{eq4} for $\alpha_1$, $\beta_1$, and $\gamma$ in the left panel and Equation \ref{eq5} for $\alpha_2$ and $\beta_2$ in the right panel.}
}
\label{fit_Sat_number_R}
\end{table}
Figure ~\ref{fig_radial} demonstrates clear gradients in the radial profiles of the color, metallicity, and age, which are consistent with the results in previous works \citep[e.g.,][]{Contini2019,Edwards2020}.
Furthermore, these gradients have pronounced dark matter halos with lower masses.
Notably, the radial profiles of the diffuse light exhibit distinct flattened distributions within approximately $\bm0.1R_{200}$, in general agreement with the observations in \cite{Yoo2021} and \cite{Montes&Trujillo2022}.
Moreover, the flattening of the color profile is more pronounced for the halos with lower masses.

The differences between the diffuse light and centrals, as well as the similarity between the diffuse light and satellites, as shown in Figure \ref{fig_properties}, can be readily explained by satellite stripping, as indicated by the gradients shown in Figure \ref{fig_radial}.
In the hierarchical clustering diagram, smaller halos form first and subsequently undergo infalling or merging to form bigger halos.
In this scenario, the substructures (or satellites) within the halos act as centrals prior to infalling, with declines in color, age, and metallicity outward along the radial profiles, corresponding to an inside-out picture for the galaxy formation \citep[e.g.,][]{Chiappini1997, vanDokkum2010, Patel2013, GonzalezDelgado2014, Tacchella2015, Carrillo2023}.
During satellite stripping caused by the tidal forces of the dark matter halo, the outer components of the satellites become separated from their central components, resulting in the formation of the diffuse light.
This process explains why the diffuse light is younger and metal poorer than that of satellites, providing an explanation for the results presented in the top and middle panels of Figure \ref{fig_properties}.


\section{Relationship between diffuse light and satellites}
\label{ICL_sat}
Here, we examine the relationship between the diffuse light and satellites.
First, we investigate the dependence of the stellar mass of the diffuse light on the total stellar mass and the number of satellites within the same dark matter halos.
Subsequently, we take advantage of the age--metallicity planes of the satellites to confirm the types of satellites that are most closely associated with the diffuse light formation.
\begin{figure*}
\centering
\includegraphics[width = 1\textwidth]{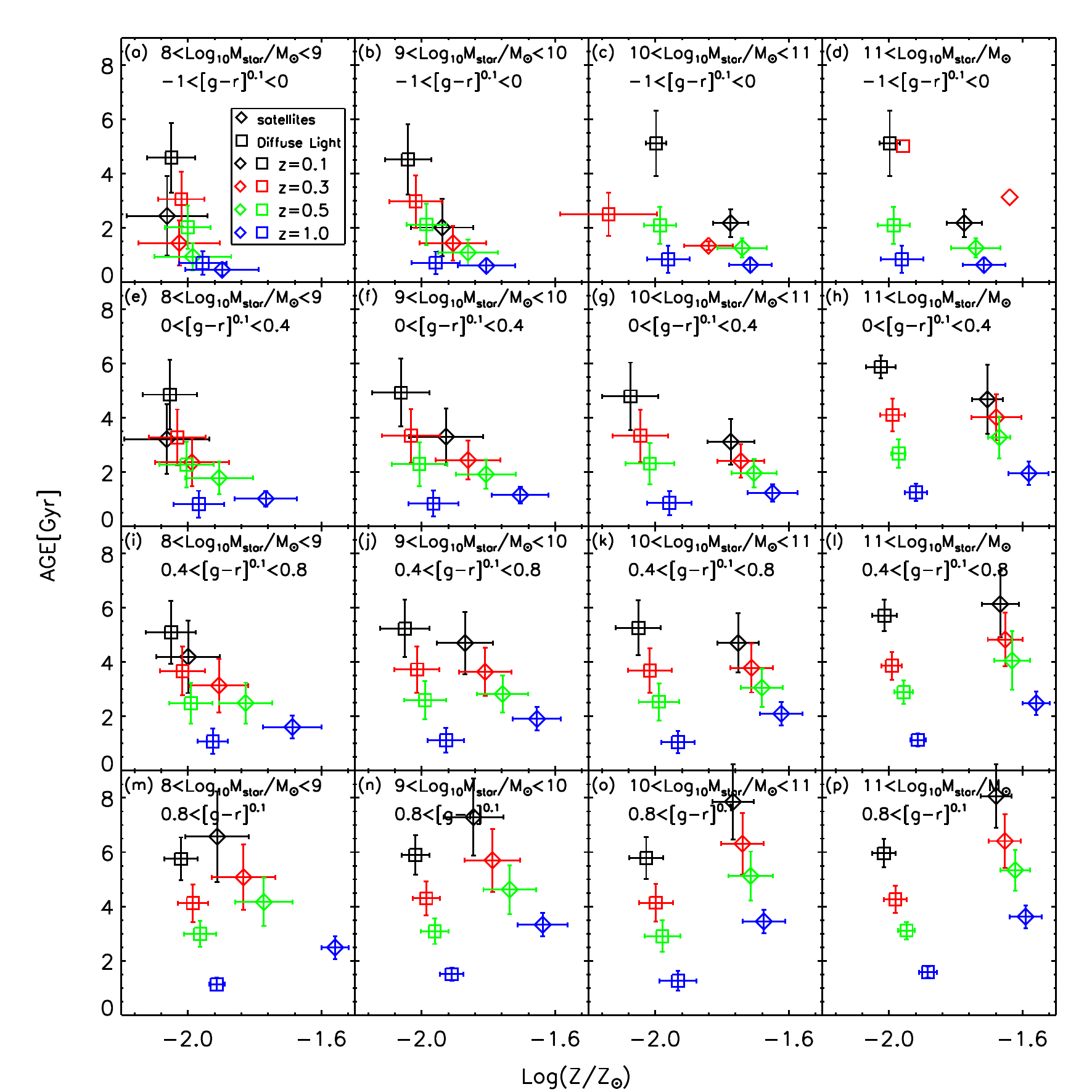}
\caption{ 
Satellite age--metallicity planes, compared to those of the diffuse light.
$Open \ diamonds$ and $open \ squares$ represent the average values of the satellite and diffuse light properties located in the same dark matter halo, respectively.
The black, red, green, and blue symbols represent the results at redshifts $z=0.1$, $0.3$, $0.5$, and $1.0$, respectively. 
The error bars represent the standard deviations.
We divide the stellar masses $\rm M_{star}$ of the satellites or color $\rm[g-r]^{0.1}$ into four bins, $\rm 8<\log_{10}M_{star}/M_{\odot}<9$, $\rm 9<\log_{10}M_{star}/M_{\odot}<10$, $\rm 10<\log_{10}M_{star}/M_{\odot}<11$, {and} $\rm 11<\log_{10}M_{star}/M_{\odot}$, from left to right, and $\rm -1<[g-r]^{0.1}<0$, $\rm 0<[g-r]^{0.1}<0.4$, $\rm 0.4<[g-r]^{0.1}<0.8$, {and} $\rm 0.8<[g-r]^{0.1}$, from top to bottom.
            }
\label{fig_Sat_properties_plane}
\end{figure*}
\subsection{Dependence of diffuse light on the stellar mass and number of satellites}\label{number_relation}
Figure \ref{fig_Sat_number_R} presents the relationship between the stellar mass of the diffuse light and two variables: the total stellar mass of the satellites (left panel) and the number of satellites (right panel) within their host halos.
The open circles represent our results at four different redshifts, distinguished by various colors, as indicated in the labels.
Note that the stellar masses of all the satellites shown in Figure \ref{fig_Sat_number_R} are greater than $10^{8}\Msun$.

As shown in the left panel of Figure \ref{fig_Sat_number_R}, we fit the relationship between the stellar mass of the diffuse light and satellites by a power-law equation, as follow:
\begin{equation}
	\label{eq4}
         y= \alpha_1 m^{\gamma} + \beta_1,
\end{equation}
where $ m=(\log_{10}(\rm M_{Sat}/\Msun))/10)$, $ y=\log_{10}(\rm M_{\rm diffuse\ light}/\Msun)$, {$\alpha_1$ is the slope, $\beta_1$ is the intercept, and $\gamma$ is the index.}

In the right panel, the fitted equation is given by 
\begin{equation}
	\label{eq5}
         y = \alpha_2 \log_{10} N + \beta_2,
\end{equation}
Where $ y=\log_{10}(\rm M_{\rm diffuse \ light}/\Msun)$, $N$ is the number of satellites in a cluster, slope $\alpha_2$ is the slope, and $\beta_2$ is the intercept.
$\rm M_{diffuse \ light}$ in {Equations} \ref{eq4} and \ref{eq5} is the stellar mass of the diffuse light.

The parameter $\gamma$ is the best fit index.
Here, we examine the dispersion of $\gamma$.
The dispersion is calculated by the standard deviation of the chi-squared error in the {$Linfit \ Routine$} of the IDL software.
In Figure \ref{sigma_check}, we illustrate the estimated dispersion of $\log_{10}{\rm M_{star}}$ of the diffuse light as a function of $\gamma$, which allows us to determine the best-fitting index in Equation \ref{eq4}.
We find that the minimum dispersion occurs at $\gamma=7.2, 7.6, 8.5$, and $9.1$ at redshifts $z=0.1, 0.3, 0.5$, and $1.0$, respectively.
Additionally, note that higher redshifts correspond to larger $\gamma$ values.
We do not present the dispersion of $\rm \log_{10} M_{star}$ of the satellites here, as it steadily decreases with increasing $\gamma$, eventually dropping to nearly 0 (around $10^{-4}$) at $\gamma=2$.
Table \ref{fit_Sat_number_R} provides the best-fitting parameters for Equations \ref{eq4} and \ref{eq5}.
Notably, the slopes $\alpha_1$ and $\alpha_2$  exhibit a decreasing trend as the redshift increases.
Conversely, $\beta_1$ and $\beta_2$ show an increase with higher redshifts.

At a given redshift (e.g., $z=1.0$, represented by the blue line in the left panel of Figure \ref{fig_Sat_number_R}), the increasing trend of the diffuse light stellar mass with {a} higher satellite mass within the range of $\rm\log_{10}M_{star}/\Msun\lesssim10$ is less steep than that within the range of $\rm\log_{10}M_{star}/\Msun\gtrsim10$.
The different colored lines in the left panel of Figure \ref{fig_Sat_number_R} demonstrate that the slopes of the increasing trend become steeper at lower redshifts within the range of $\rm\log_{10}M_{star}/\Msun\lesssim10$, and vice versa within the range of $\rm\log_{10}M_{star} /\Msun\gtrsim10$.
In the right panel of Figure \ref{fig_Sat_number_R}, the data points at each redshift exhibit a close-to-linear relationship.
As shown in the inset of the right panel of Figure \ref{fig_Sat_number_R}, the stellar mass of the diffuse light exhibits a steep increase with the number of satellites (N) up to approximately 50 satellites, after which the trend becomes nearly flat for N $\gtrsim$ 100.
In conclusion, the diffuse light components tend to have higher average masses within host halos that contain a greater number of satellites or more massive satellites.
 
From the left panel of Figure \ref{fig_Sat_number_R}, It seems that the mass of the diffuse light increases with increasing redshift. 
We have checked the relative mass or light fractions of the diffuse light compared to the total stellar masses within the host halos, and find that these fractions are about 21\%, 27\%, 33\%, and 46\% at redshifts z = 0.1, 0.3, 0.5, and 1.0 respectively. 
This finding agrees with the results presented in Figure 6 of \cite{Tang2018} but contradicts with the diffuse light evolution described in \cite{Burke2015} who introduced an evolving factor for the SBL, i.e.,  $\bm 2.5\log(1+z)^4$, meaning that the SBL at higher redshift will be fainter. 
Similarly, as has been done in \cite{Tang2018}, if we also made such a cosmological correction on the SBL, the stellar mass fractions are about 17\%, 16\%, 13\%, and 12\% at redshift z = 0.1, 0.3, 0.5, and 1.0 respectively, in good agrees with those of previous works \citep[e.g.,][]{Rudick2011, Burke2015, Tang2018, Montes&Trujillo2018}, with decreasing diffuse light fraction to higher redshifts. 
Otherwise, if not applying such SBL correction, the predictions of increasing diffuse light fraction with increasing redshift are primarily caused by observational effects. 
At high redshift, many satellites with low surface brightness are smoothing out and missed in the projecting image by the SBLSP method, as the galaxies must be brighter than the SBL ($\bm26.5 \ mag/arcsec^2$ in this paper).  
Therefore, it turns out that the number of satellites will be smaller at higher redshift, as shown in the right panel of Figure \ref{fig_Sat_number_R}. 
The stellar mass of these missing satellites will be then classified as the diffuse light, causing the larger total mass of the diffuse light at higher redshift. 
Nevertheless, the contradiction with traditional expectation should be resolved and further investigations are required, although it is not the primary focus of this paper.

\subsection{Age--metallicity planes of diffuse light and satellites}\label{property_plane}

Informed by the study in \cite{Contini2019}, we aim to gain further insights into the crucial types of satellites that contribute significantly to the diffuse light formation.
To that end, we investigate the relationship between the intrinsic properties of the diffuse light and satellites, considering their redshift evolution.
In Figure \ref{fig_Sat_properties_plane}, we present the age--metallicity planes of the satellites, represented by open diamonds,  along with the corresponding average properties of the diffuse light,    indicated by open squares, allowing for direct comparisons.

Note that we employ the age--metallicity plane, which differs from the color--metallicity plane utilized in \cite{Contini2019}.
As shown in Figures \ref{fig_properties} and \ref{fig_distribution}, the color ($\rm[g-r]^{0.1}$) distribution of the diffuse light exhibits a closer resemblance to that of the centrals than that of the satellites, particularly at lower redshifts.  
This indicates a potential coevolution of the diffuse light and centrals in terms of color.
However, discernible variations are found in the age and metallicity distributions of the diffuse light.
The peaks of the age distribution of the diffuse light assemble with those of the satellites, while the peaks of the metallicity distribution of the diffuse light are lower than those of the satellites.
Consequently, by employing the age--metallicity plane of the satellites, we effectively mitigate scattering effects and gain a clearer understanding of the intrinsic properties. 
Similarly to \cite{Contini2019}, we classify the satellites into four stellar mass bins to identify which types of satellites exhibit the most similar properties of the diffuse light.
Additionally, we categorize the satellites into four color bins, enabling a comprehensive exploration of the satellite--diffuse light relationship.

From the overall view of Figure \ref{fig_Sat_properties_plane}, it is found that the properties of the satellites and diffuse light are located in close regions in the age--metallicity planes only in panels (a) and (b) for all four redshifts.
From the right panels to the left panels, the plotted points for the diffuse light and satellites move closer to each other in the age--metallicity planes at four redshifts, for all the color bins.
From the top panel to the bottom panel, the bluer diffuse light and satellites become more similar.
However, in panels (d), (h), (l), and (p) (most massive mass bins) and in panel (o), no clear correlation is found between the diffuse light and satellites.

It is important to determine which kinds of satellites are related to the formation of the diffuse light at high or low redshifts.
In our results, at redshift $z=1.0$, the satellites with stellar masses within the range of $\rm 8<\log_{10}M_{star}/M_{\odot}<10$ and colors within the range of $\rm-1<[g-r]^{0.1}<0$ (panels (a) and (b)) are the primary sources of the formation of the diffuse light.
At lower redshifts, the redder satellites (panels (e), (f), and (i)) are related to the build-up of the diffuse light.
However, we cannot confirm the relationships between the smallest and reddest satellite (panel (m)) and the formation of the diffuse light.


\section{Discussion and Conclusions}\label{conclusions}

In this study, we utilize the TNG-100 simulation of IllustrisTNG to investigate the age, metallicity, and color $\rm[g-r]^{0.1}$ of the diffuse light, as well as their relationships with the centrals and satellites.
To distinguish between the diffuse light and galaxies, we employ an improved surface brightness level segmentation procedure called SBLSP.
This procedure is based on an iterative surface brightness level cutting in projected mock images generated by converting the stellar particles from TNG100-1 snapshots into seeing-convolved pixelated images.
To obtain realistic observational conditions and reproduce imaging-like data, we consider Moffat PSF, dust extinction, and redshift dimming effects.
Furthermore, to reduce resolution effects in the simulations, we focus on the dark matter halos with total mass $\rm M_{200}$ greater than $\rm 10^{12}M_{\odot}$ and the stellar mass of galaxies $\rm M_{star}$ greater than $\rm 10^{8}M_{\odot}$.

We perform a comprehensive analysis of various properties of the diffuse light, including the age, metallicity, color $\rm [g-r]^{0.1}$, and stellar mass. 
We then examine the number distributions of these properties and investigate their dependence on the dark matter halo mass $\rm M_{200}$. 
Additionally, we compare these distributions with those of the centrals and satellites to gain a better understanding of the formation processes.
To further explore the nature of the diffuse light, we analyze their radial profiles, which allows us to validate the presence of gradients of properties such as the color, metallicity, and age, which are consistent with previous studies.
Furthermore, we examine the relationship between the stellar mass of the diffuse light and the stellar mass or number of satellites within the same dark matter halos.
Finally, we explore the age--metallicity planes for the satellites and diffuse light to gain insights into their intrinsic properties and evolution.
The results are summarized as follows.
\begin{itemize}
\item[1.] 
The age and metallicity of the diffuse light components resemble those of the satellites, and both are distinct from those of the centrals.
However, as the redshift decreases, the color of the diffuse light approaches that of the centrals. 
These findings indicate that the diffuse light components are associated with satellite galaxies, and they gradually turn red by passive evolution with the centrals (coevolution). 
\item[2.] 
Consistent with recent observational findings, the age, metallicity, and color radial profiles of the diffuse light appear flat in the inner region but drop sharply in the outer region.
This finding provides a clue for the build-up of the diffuse light by mixed mechanisms of the major merger and tidal stripping.
\item[3.]
The stellar mass of the diffuse light correlates strongly with the total stellar mass and number of satellites in the host halo, and these relationships can be modeled using a power-law and logarithmic function, respectively.
\item[4.]
Satellites with stellar masses within the range of $\rm8<\log_{10}M_{star}/M_{\odot}<10$ and colors within the range of $\rm-1<[g-r]^{0.1}<0$ are found to be the primary source of the diffuse light.
From high to low redshifts, an increasing number of redder satellites participate in the formation of the diffuse light, whereas the most massive and reddest satellites contribute little to the diffuse light formation.
\end{itemize}

Previous studies \citep[e.g.,][]{Montes&Trujillo2014, Morishita2017,Contini2019} have commonly utilized diffuse light color diagrams to investigate the satellites contributing to the diffuse light.
However, our study reveals a coevolutionary relationship between the color of the diffuse light and centrals, as well as a wide distribution of the diffuse light color.
We caution readers to avoid relying solely on diffuse light color diagrams, as they may introduce scatter and potentially lead to spurious conclusions.
In Figure \ref{fig_Sat_number_R}, we also suggest a novel approach for determining the stellar mass or number of satellites in dark mark halos.

Notably, our result slightly differs from that in \cite{Contini2019}, which is that the low-mass galaxies, with $\rm 9<log_{10}M_{star}M_{\odot} <10$, are the most important contributors {of the diffuse light} at high redshifts, but that massive contents are contributed by intermediate/massive galaxies, with $\rm10<log_{10}M_{star}/M_{\odot} <11$, at lower redshifts.
The stellar masses of the satellites relative to the diffuse light formation in \cite{Contini2019} are roughly one order of magnitude higher than our results.
The main reason is that the mock galaxies defined by our method are typically less massive than those found by the traditional substructure extraction algorithm \citep{Tang2021}, because the method distinguishes the diffuse light located around the galaxies.
Otherwise, our results confirm that the satellites that formed the diffuse light are blue galaxies and show color--redshift dependence.

By applying cosmological hydrodynamical simulations with high resolution, we can improve our theoretical understanding of the spatial distribution of the diffuse light, as discussed in Section 5.
We find a flattened distribution of the diffuse properties in the innermost region and a significant drop in the outer region.
However, only color and metallicity gradients have been found in previous observational and theoretical works \citep[e.g.,][]{Montes&Trujillo2018, Contini2019}.
The platform in the innermost region of the diffuse light properties, such as the color, has been confirmed in recent observations \citep[e.g.,][]{Yoo2021, Montes&Trujillo2022}.

Furthermore, the relationship between the stellar mass of the diffuse light and the number of satellites serves as an effective indicator of the interaction efficiency within groups and clusters of galaxies, considering that the diffuse light arises from merging and tidal stripping processes \citep[e.g.,][]{Montes2019}. 
Consistent with previous studies \citep[e.g.,][]{Rudick2006, Montes&Trujillo2018}, our findings demonstrate that groups and clusters of galaxies with a higher abundance of the diffuse light tend to include a greater number of satellites, enabling more frequent satellite interactions.
Conversely, groups and clusters of galaxies with smaller numbers of satellites are less efficient in terms of interactions, resulting in slower increases in the stellar mass of the diffuse light.

Overall, our investigations provide a crucial insight into the formation and evolution of the diffuse light.
To gain a deeper understanding of the specific mechanisms involved in the formation and evolution of the diffuse light, in the future, we plan to trace the spatial distribution of stellar particles to elucidate their origins and the processes through which they become diffuse light components.


\section*{Acknowledgments}
The authors thank the anonymous referee for the useful suggestions and the Illustris and IllustrisTNG projects for providing simulation data.
We acknowledge support from the NSFC grant (No. 12073089, No. 12003079) and the National Key Program for Science and Technology Research and Development (No. 2017YFB0203300).
L.T. is supported by Natural Science Foundation of Sichuan Province (No. 2022NSFSC1842), the Fundamental Research Funds of China West Normal University (No. 21E029), and the Sichuan Youth Science and Technology Innovation Research Team (21CXTD0038). 
Y.W. is supported by The Major Key Project of PCL.
J. L. is supported by an NSFC grant (No. 12273027).
The analysis carried out in this work is done at the Kunlun HPC facility of the School of Physics and Astronomy, Sun Yat-Sen University.
We thank LetPub (www.letpub.com) for its linguistic assistance during the preparation of this manuscript.


\bibliographystyle{apj}
\bibliography{main}

\end{CJK*}
\end{document}